\documentclass{article}

 \usepackage[preprint]{neurips_2026}

\usepackage[utf8]{inputenc} 
\usepackage[T1]{fontenc}    
\usepackage{hyperref}       
\usepackage{url}            
\usepackage{booktabs}       
\usepackage{amsfonts}       
\usepackage{nicefrac}       
\usepackage{microtype}      
\usepackage{xcolor}         
\usepackage[pdftex]{graphicx}
\usepackage{comment}
\usepackage{fancyvrb}
\usepackage{float}
\usepackage[T1]{fontenc}
\usepackage{tikz}
\usetikzlibrary{arrows.meta, positioning, shapes.geometric}

\title{STEM-Bank-26: An Open Dataset of Analytical Scanning Transmission Electron Microscopy Images and Spectra}

\author{%
  George G. Hollyer\\
  University of Pennsylvania\\
  Philadelphia, PA 19104\\
  \texttt{ghollyer@seas.upenn.edu} \\
  \And
  Eric A. Stach\thanks{Corresponding author: stach@seas.upenn.edu}\\  
  University of Pennsylvania\\
  Philadelphia, PA 19104\\
  \texttt{stach@seas.upenn.edu} \\
}

\begin{document}

\maketitle

\begin{abstract}
 Scanning transmission electron microscopy (STEM) is a versatile, multi-modal method for studying materials at the nanoscale. It integrates multiple signal sources to provide atomic resolution imaging alongside spectroscopy for elemental composition and electronic structure. Despite its utility, few datasets spanning multiple years of data have been published. We introduce STEM-Bank-26, a 929 GB open dataset of 37,921 STEM images that spans 1,095 material samples across 1,499 sessions and includes energy-dispersive spectroscopy (EDS) or electron-energy loss spectroscopy (EELS) data for 65.5\% of sessions. We extract microscope control trajectories by comparing metadata between sequential images and provide tools for building similar datasets from archival data. To our knowledge, this is the largest openly available dataset that integrates experimental STEM, EDS and EELS data across a broad distribution of samples and sessions.
\end{abstract}

\section{Introduction}

Scanning Transmission Electron Microscopy (STEM) is an imaging technique capable of directly imaging materials with resolution around or under one angstrom (Å) or 10$^{-10}$ m \citep{pennycookScanningTransmissionElectron2011}.  STEM is commonly paired with the spectroscopic techniques energy-dispersive spectroscopy (EDS) and electron energy-loss spectroscopy (EELS) that extract elemental and electronic structure information from electron-sample interactions. Pairing atomic resolution imaging with spectroscopic information is uniquely valuable for studying structure-property relationships in areas such as energy storage, heterogeneous catalysis, nanomaterial synthesis, semiconductor devices and environmental remediation. 

Automated TEM has been a longstanding area of interest and more recently has become a priority area due to its potential utility in closed-loop autonomous materials research \citep{liAITransmissionElectron2025, jamaliThinkingMicroscopesAgentic2026, kalininHumanintheLoopFutureMachine2024}.  However, efforts to build ``AI scientists'' or TEM operators are hampered by the difficulty of training AI models to carefully explore and adapt to new information.

Electron microscopy is an inherently multi-turn, exploratory task with a large space of possible actions. Operators rely on their judgment and experience to identify areas of interest, change imaging conditions and employ spectroscopic techniques to extract useful information from the sample. Existing automation efforts have focused on purpose-trained models for a specific task or are confined to text-only assistants rather than creating generalist models with vision capabilities \citep{wallTEMAgentEnhancing2026}. In the broader robotics literature, recent advances in building integrated vision-language-action (VLA) models have shown the importance of curating human trajectories for imitation learning \citep{embodimentcollaboration2025openxembodimentroboticlearning}. Microscopy has no equivalent corpus of recorded trajectories but does have large archival datasets with attached metadata for each acquisition. Given the expense of microscope time and small number of instruments, reconstructing partial trajectories from these datasets is the only feasible way of building a large trajectory corpus in the near term. 

Several recent commentaries have identified a tendency to only publish small subsets of collected microscopy data \citep{ivanovaLostDataElectron2025,kalininWeReallyNeed2026a}. A handful of representative images are typically sufficient for the scientific conclusions of a publication and file sizes for a single session easily reach into the gigabytes, posing a logistical challenge. With increased interest in automation and deep learning methods, the need for open datasets is growing and the microscopy community is rethinking these norms.

This work aims to publish a large corpus of analytical STEM data and show how partial control trajectories can be extracted via metadata analysis. Alongside the dataset, we provide scripts to automatically analyze and build trajectory datasets from existing archives of common Gatan DigitalMicrograph files so others can easily convert their archival datasets to trajectories. 

\subsection{Prior Work}

Several large, openly available experimental TEM/STEM datasets for machine learning have been published including the Warwick electron microscopy dataset  \citep{edeWarwickElectronMicroscopy2020}, HRTEM images of nanoparticles collected at Lawrence Berkeley National Laboratory \citep{sytwuHRTEMImagesGeneralization2023} and a more recent dataset of STEM images from Empa Dübendorf \citep{channingContrastiveImageMetadataPreTraining2026}. These collections demonstrate the utility of large-scale experimental microscopy data for machine learning but do not contain spectroscopic data. Therefore, they do not reflect the multi-modal capabilities of analytical electron microscopy. Comparable session/sample/trajectory counts for these datasets cannot be calculated due to their partial metadata. 

Some experimental spectroscopic data exists in the form of reference databases and publications \citep{ewelsCompleteOverhaulElectron2016, ahnEELSAtlasReference1983} but they contain no TEM/STEM images from the acquisition session and are only single spectra instead of SI. Simulated EELS datasets have been created for machine learning \citep{annysDeepLearningAutomated2023a}, but do not capture the full variation of experimental data. To our knowledge there is no comparably large, openly available dataset that integrates experimental STEM, EDS, and EELS data across a broad distribution of samples and sessions. 

\begin{table}[ht]
    \centering
\caption{Comparison of Large Open Experimental Datasets for Machine Learning}
\label{tab:datasetComparison}
    \begin{tabular}{ll|rrrr}
 &  &  \multicolumn{4}{c}{No. of Items}\\\toprule
 Publication& Size & TEM&STEM&EDS&EELS\\\midrule
         
         Ede 2020& 429 GB$^*$ & \textbf{17,266}&  19,769&  -& -\\
         
         Channing et al. 2026& 128 GB & -&  7,330&  -& -\\
         Sytwu et al. 2023& 33.7 GB & 407&  -&  -& -\\
 Ewels et al. 2016& 7.82 MB &  -& -& -&290$^\dag$\\
         This work& 929 GB & -&  \textbf{37,921}&  \textbf{4,996}& \textbf{7,707}\\

    \end{tabular}
    \begin{minipage} {\linewidth}
        \small
        \vspace{0.2cm}
        \textit{$^*$Computed from full resolution TEM and STEM images only}
        
        \textit{$^\dag$Individual spectra, not spectrum images}
    \end{minipage}
    
\end{table}
\begin{figure}
    \centering
    \includegraphics[width=1\linewidth]{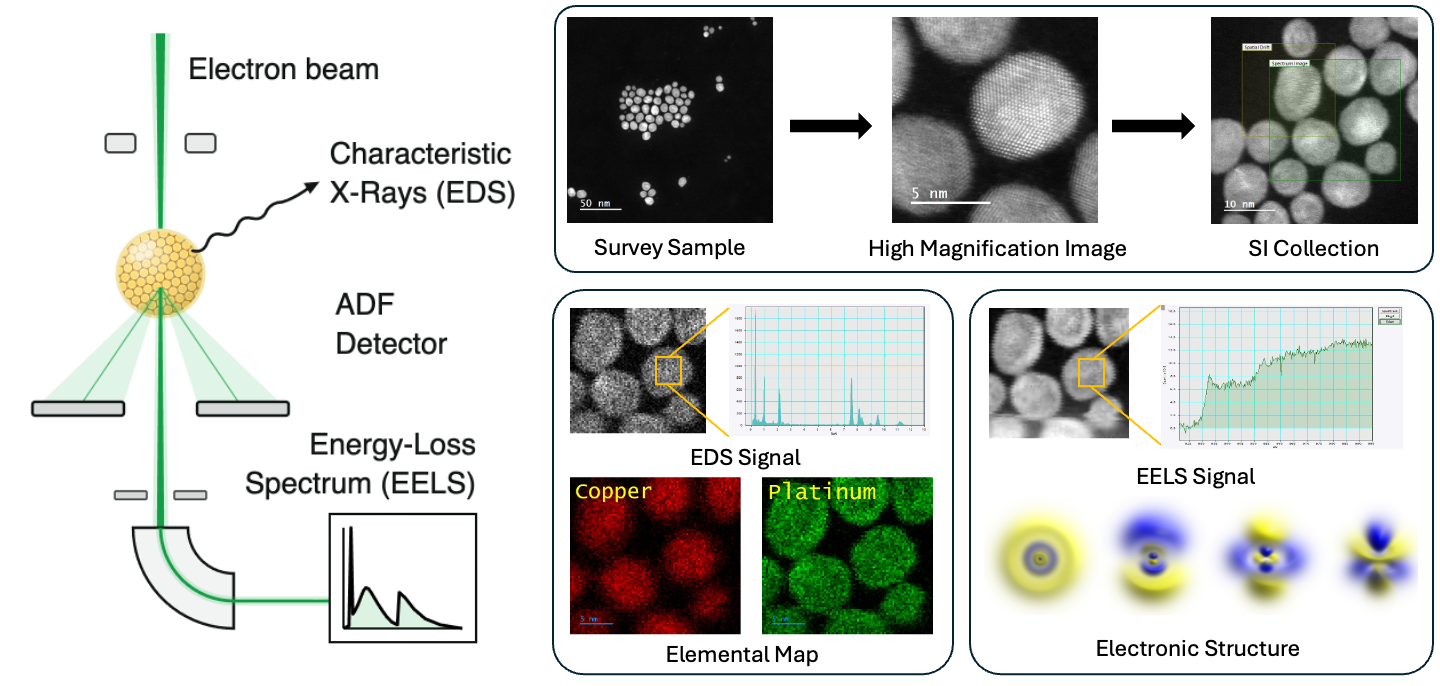}

    \caption{(Left) Multi-modal data collection in STEM from scattered electrons (ADF-STEM), characteristic X-rays (EDS) and electron energy loss (EELS). (Right) Example of a typical analytical STEM trajectory in the dataset beginning with surveying the sample and imaging at high magnification before choosing an area to collect EDS and EELS spectrum images. EDS and EELS data are used to map copper and platinum morphology, analyze the oxidation state of copper and estimate the elemental composition.}
    \label{fig:paperOverview}
\end{figure}

\section{Methods}

We have aggregated a dataset of 1,499 microscope sessions taken between 2019 and 2024 on a JEOL NEOARM STEM equipped with EDS detectors and a Gatan GIF Quantum EELS spectrometer. Data is in the form of Gatan DigitalMicrograph (.dm4/.dm3) files free of post-session processing. Files retain the metadata available at time of acquisition including stage position, timestamp, magnification, dwell time, camera length, calibrated intensity and spectroscopy parameters. 

The microscope collects three forms of signals. High-resolution images are formed from the high-angle scatter of electrons by the sample onto an annular dark-field (ADF) detector. Areas with no sample appear dark and intensity increases with the atomic number of the scattering atoms in a modality called Z-contrast. The EDS and EELS detectors can simultaneously collect spectroscopic information. The EDS system captures X-ray fluorescence and compares the spectrum to known element characteristic X-ray energies. The EELS spectrometer measures the energy distribution of electrons exiting the sample and produces a spectrum of energy losses. This serves as a probe of sample electronic structure and can be analyzed to determine chemical information such as the oxidation state of the scattering element. By collecting spectra at different probe positions, the microscope builds a "spectrum image" (SI) comprised of two spatial dimensions and one spectral dimension \citep{fultzTransmissionElectronMicroscopy2013}\citep{williamsTransmissionElectronMicroscopy2008}. 

For analysis, we build an SQLite database of image metadata from which we reconstruct sessions and trajectories. Sessions are determined based on unique sample-day pairs. Trajectories are assembled by logging the metadata changes between consecutive images and SI in a session within thresholds of stage movement and time between samples. By default, we choose thresholds of 1 hour between images or 50 microns of stage position. The assumption being, if the operator moves to an unseen area of the sample or waits an unusually long time between two images then the context has been severed. This is an arbitrary threshold and because the full metadata is preserved, users could build trajectories with different thresholds or introduce entirely new criteria.

\section{Contributed Data}

\begin{figure}
    \centering
    \includegraphics[width=0.75\linewidth]{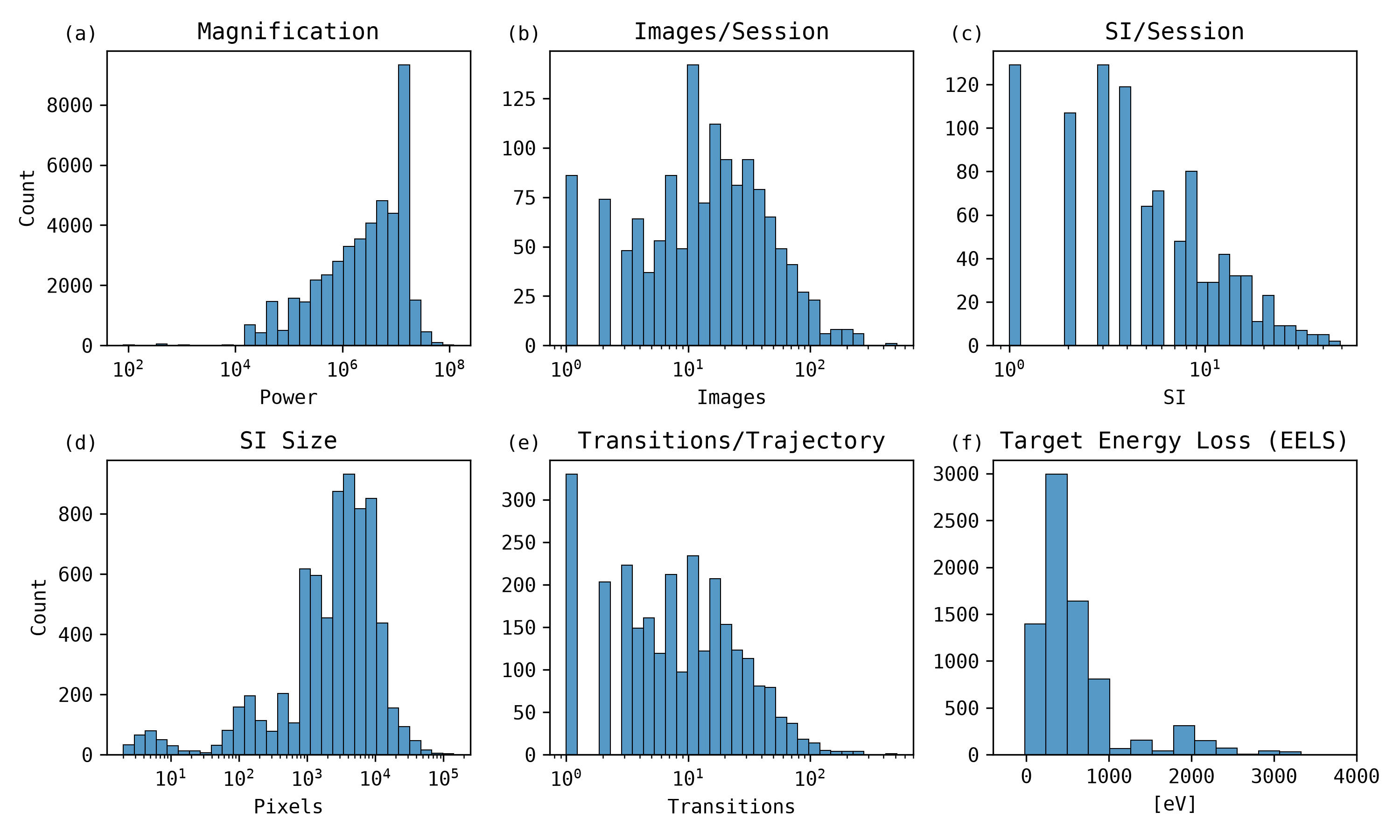}
    \caption{Distribution of magnification, images and SI per session and trajectory length}
    \label{fig:statistics}
\end{figure}

\begin{table}[ht]
\centering
\caption{Session and trajectory statistics}
\begin{tabular}{lrrrrr}
\hline
Metric & $N$ & Min & Mean & Median & Max \\
\hline
Images per session & 1499& 0 & 27.85 & 13& 516\\
SI acquisitions per session & 1499& 0 & 4.63& 2 & 49\\
Transitions per trajectory & 3191& 0 & 13.12& 6 & 515\\
\hline
\end{tabular}
\end{table}

The dataset is comprised of 37,921 STEM images, 4,996 EDS spectrum images and 7,707 EELS spectrum images in the form of Gatan DigitalMicrograph files. Through filename and metadata analysis we infer 1,095 unique samples across 1,499 sessions. To not overstate the number of samples examined, textual analysis is performed to de-duplicate samples by identifying condition words such as "old", "repeat" and "used" \ref{fig:processing}. Elements present are inferred from sample and file names after filtering for confounding terms similar to chemical formulas such as "NP" for nanoparticle and "SEI" for solid-electrolyte interphase. Very unlikely elements such as noble gases and transuranics are excluded and the results are plotted in \ref{fig:elements}. A set of 100 randomly chosen samples with 214 inferred element labels was audited and the element label F1 score was calculated at 93.3\% \ref{elementLabelMetrics} 

For all SI, a survey STEM image used to specify the collection area is saved. Therefore, image-SI pairs are available for all SI. These survey images are not included in the overall count of STEM images as they often have the same field of view as a standalone STEM image. 

The dataset is predominantly comprised of heterogeneous catalyst and lithium ion battery samples with a long tail of samples for other applications. For EDS and EELS the number of spectrum images and unique samples exceeds that of current public databases but has a narrower elemental coverage\citep{ewelsCompleteOverhaulElectron2016, ahnEELSAtlasReference1983}.

\begin{table}
    \centering
\caption{Session-level spectrum image coverage}
\label{siCoverageTable}
    \begin{tabular}{lrr}\toprule
         Category&  Sessions& Percent\\\midrule
 None& 517&34.48\%\\
         EDS Only&  341& 22.75\%\\
         EELS Only&  161& 10.74\%\\
 EDS and EELS& 480&32.02\%\\
 \hline
 Total sessions& 1499&100\%\\ 
 \hline
    \end{tabular}
    
\end{table}
There is some variation of image size. STEM images are predominantly of dimension 1024x1024 (90.8\%) with 512x512 (6.1\%) and 2048x2048 (0.97\%) forming the majority of the remaining images. EDS and EELS spectrum images are of variable user-defined dimensions and are not constrained to be square. 63.9\% of spectrum images contain between 10$^3$ and  10$^4$ pixels.

Sessions have a variety of length with 39.43\% of sessions having <10 images, 57.57\% between 10 and 100 images and 3.00\% having >100 images. Some sessions do not contain images and consist only of SI data meaning they do not build trajectories. However, their contribution to the dataset is minor with just 80 such sessions (5.34$\%$). 

Due to the underlying data, this work has some inherent limitations. First, trajectories are "snapshots" of microscope state and do not capture the live search scans or full operator inputs between them. For instance, the stage may have been moved to inspect several areas where no images were taken. This could bias the dataset towards specimen areas deemed high enough quality for a full-resolution capture. Second, lens parameters for the microscope and aberration corrector are not present in metadata from this period. Therefore, this dataset is of limited utility for microscope alignment and tuning tasks. Metadata is presented as it exists in the underlying file and contains some missing/duplicate fields due to incomplete integration of DigitalMicrograph with the microscope. \ref{exampleMetadata}

We intend for this work to be a usable training corpus. Lists of samples, sessions and trajectories are constructed as SQLite databases that can be queried by dataloaders. The full metadata database is provided for the convenience of researchers who wish to only work with a subset of the data or group it according to their needs without having to download all files. We also provide the scripts used to assemble these databases to help researchers add or publish microscope data in their possession. Building databases for the full 929 GB dataset took approximately 45 minutes on a consumer laptop.

\section{Conclusion}

We introduce a large dataset of STEM images and spectra collected on an aberration-corrected analytical STEM over a period of five years. Metadata analysis shows wide coverage of samples, elements, magnification, session length and target energy-loss. Microscope control trajectories are produced by analyzing metadata transitions between sequential images in a session. These trajectories contain the context for choosing to acquire an SI in a particular area and the settings used for acquisition. Statistical analysis shows images and spectra are widely distributed over the sessions and samples, with 57.57 $\%$ of sessions between 10 and 100 images in length and 65.51\% having some form of spectral data. To our knowledge, this work is the largest published dataset of integrated STEM, EDS and EELS data. 

\section{Availability}

The full dataset and codebase is available at https://huggingface.co/buckets/ghollyer/STEM-Bank-26 and at the Materials Data Facility (https://www.materialsdatafacility.org/) under a CC BY 4.0 license. 

The codebase is additionally available at https://github.com/ghollyer/STEM-Bank-26

\begin{ack}
This work was carried out in part at the Singh Center for Nanotechnology, which is supported by the NSF National Nanotechnology Coordinated Infrastructure Program under grant NNCI-2025608. Additional support to the Nanoscale Characterization Facility at the Singh Center for Nanotechnology is provided by the NSF through the University of Pennsylvania Materials Research Science and Engineering Center (MRSEC) (DMR-2309043). This work is supported by the Interdisciplinary Training in Data Driven Soft Materials Research and Science Policy (D2SMRSP), NSF Award \# 2152205.

The authors thank Georgia Channing for her feedback on this manuscript

"Atomic-orbital-clouds spdf m0" by Geek3 is licensed under CC-BY-SA 4.0 https://commons.wikimedia.org/wiki/File:Atomic-orbital-clouds\_spdf\_m0.png

\end{ack}

\newpage

\bibliographystyle{unsrtnat}
\bibliography{reference}
\newpage

\appendix
\section{Appendix}
\setcounter{figure}{0}
\setcounter{table}{0}
\renewcommand{\thefigure}{A.\arabic{figure}}
\renewcommand{\thetable}{A.\arabic{table}}

\subsection{Sample Images}
\begin{figure}[H]
    \centering
    \includegraphics[width=0.9\linewidth]{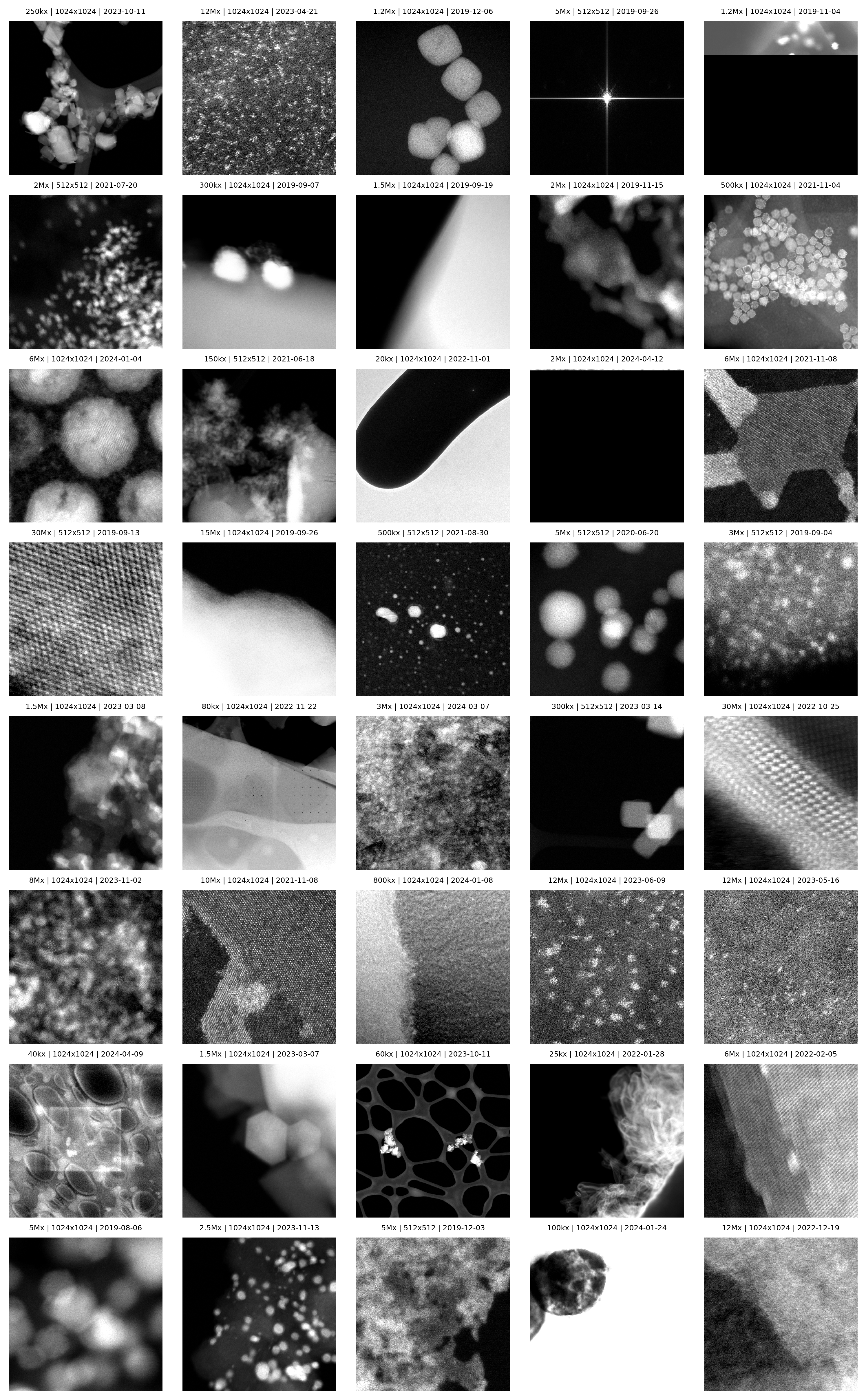}
    \caption{40 randomly sampled ADF-STEM images}
    \label{fig:placeholder}
\end{figure}

\subsection{Elemental Distribution}

\begin{figure}[H]
    \centering
    \includegraphics[width=1\linewidth]{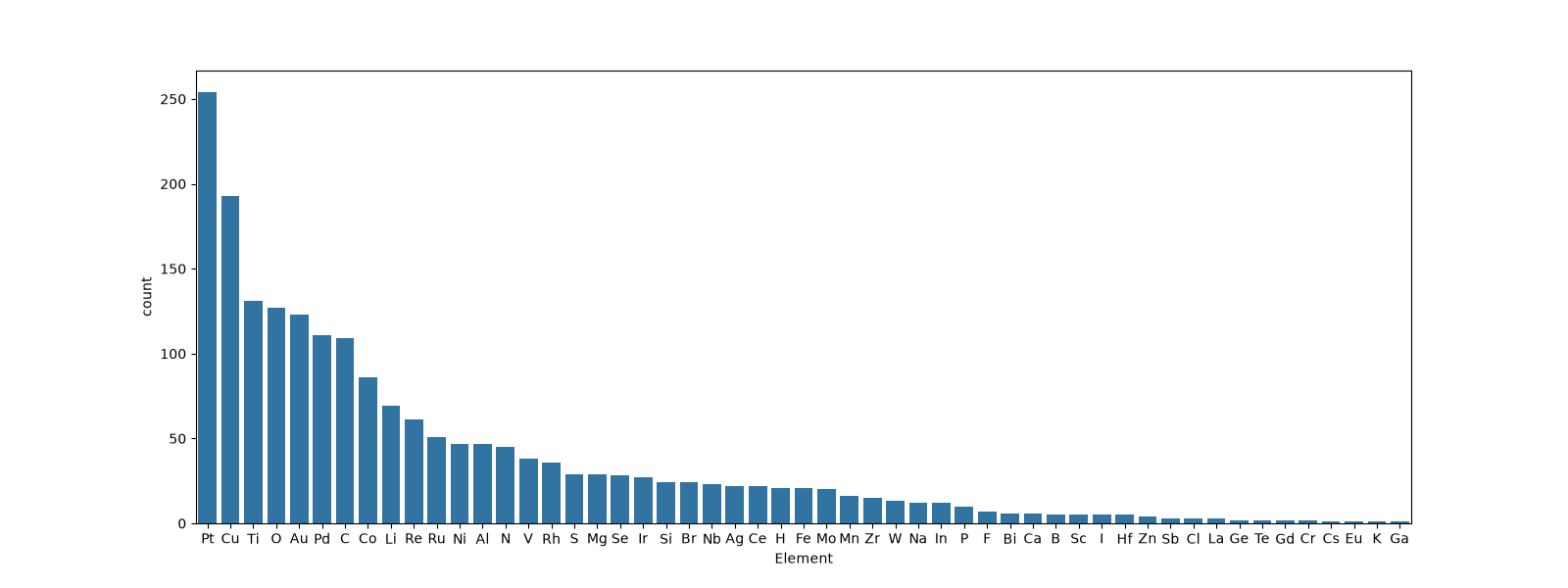}
    \caption{Number of material samples containing each element inferred from metadata}
    \label{fig:elements}
\end{figure}
A random set 100 samples with inferred element labels was manually audited. Recall, precision and F1 scores were computed at both the element label and sample level.

\begin{table}[ht]
    \centering
    \caption{Accuracy Metrics of Element Labels}
    \begin{tabular}{cccc}\toprule
         Metric Level&  Precision&  Recall&  F1\\\midrule
         Element&  94.9\%&  91.9\%&  93.3\%\\
         Sample&  93.3\%&  94.1\%&  89.7\%\\ \bottomrule
    \end{tabular}
    \label{elementLabelMetrics}
\end{table}

\subsection{Metadata Processing}
\begin{figure}[H]
    \centering
    \includegraphics[width=0.8\linewidth]{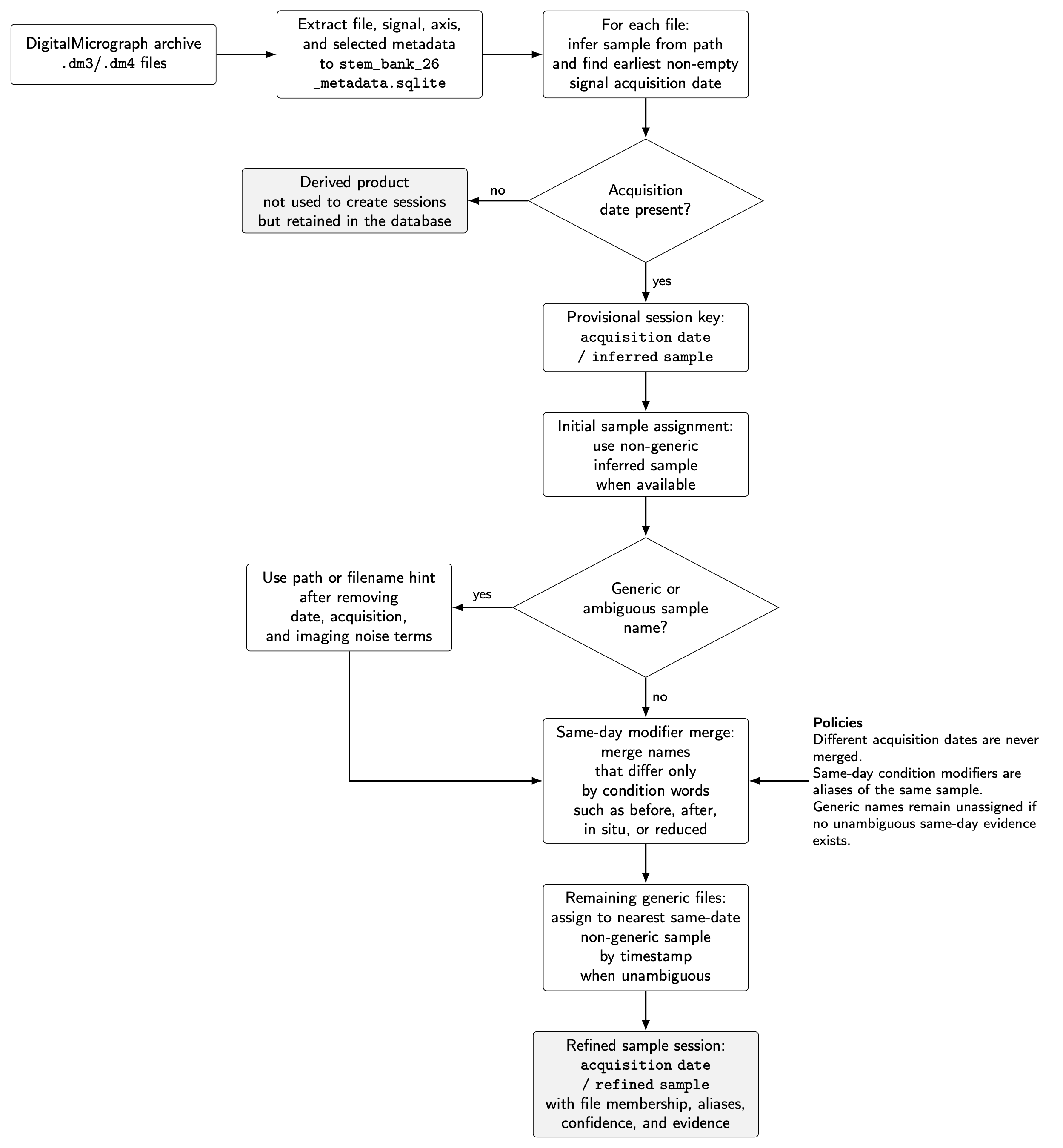}
    \caption{Processing and classification scheme for building session and sample databases}
    \label{fig:processing}
\end{figure}

\newpage
\subsection{Example Metadata}
\label{exampleMetadata}
\begin{Verbatim}[frame=single, fontsize=\small, framesep=4pt]
Example metadata entry for an ADF-STEM image
{
  "image_id": 1,
  "session_id": 1,
  "session_key": "2019-01-16/Vohs group",
  "trajectory_id": 1,
  "trajectory_key": "2019-01-16/Vohs group/trajectory_0001",
  "trajectory_sequence_index": 1,
  "source_file_id": 87984,
  "signal_id": 74724,
  "path": "Vohs group/JEOL  image 1 0014.dm3",
  "title": "JEOL  image 1 0014",
  "acquisition_date": "2019-01-16",
  "acquisition_time": "14:35:02",
  "acquisition_timestamp": "2019-01-16T14:35:02",
  "sequence_index": 1,
  "data_shape": "768x768",
  "data_dtype": "uint16",
  "stage_x_text": "-19.3760234375",
  "stage_y_text": "-20.376388671875",
  "stage_z_text": "21.140103515625",
  "stage_tilt_alpha_text": "0.0694548562169075",
  "stage_tilt_beta_text": "-0.030053284019231796",
  "magnification_text": "",
  "beam_energy_text": "",
  "beam_current_text": "",
  "camera_length_text": "",
  "dwell_time_text": "",
  "field_of_view_um_text": "0.1137203125",
  "probe_current_na_text": "0.0",
  "probe_size_nm_text": "0.0",
  "formatted_magnification_text": "3Mx",
  "formatted_voltage_text": "80kV",
  "stage_x_number": -19.3760234375,
  "stage_y_number": -20.376388671875,
  "stage_z_number": 21.140103515625,
  "stage_tilt_alpha_number": 0.0694548562169075,
  "stage_tilt_beta_number": -0.030053284019231796,
  "magnification_number": null,
  "beam_energy_number": null,
  "beam_current_number": null,
  "camera_length_number": null,
  "dwell_time_number": null,
  "field_of_view_um_number": 0.1137203125,
  "probe_current_na_number": 0.0,
  "probe_size_nm_number": 0.0,
  "formatted_magnification_number": null,
  "formatted_voltage_number": null,
  "focus_text": "",
  "defocus_text": "",
  "objective_lens_current_text": "",
  "condenser_lens_current_text": "",
  "projector_lens_current_text": "",
  "stigmation_x_text": "",
  "stigmation_y_text": ""
}
\end{Verbatim}


\end{document}